\documentclass[prd,superscriptaddress,showpacs,showkeys]{revtex4}
\usepackage{textcomp}
\usepackage{amssymb}
\usepackage{eurosym}
\usepackage{amsmath}
\usepackage{epsfig}
\usepackage{graphics}
\usepackage{color}
\usepackage{graphicx}
\usepackage[font={footnotesize,it}]{caption}
\usepackage[colorlinks=true,
            linkcolor=red,
            urlcolor=blue,
            citecolor=green]{hyperref}

\begin{document}

\title{Quantum Tunneling and Quasinormal Modes in the Spacetime of the
Alcubierre Warp Drive}

\author{Kimet Jusufi}
\email{kimet.jusufi@unite.edu.mk}
\affiliation{Physics Department, State University of Tetovo, Ilinden Street nn, 1200,
Tetovo, Macedonia} 
\affiliation{Institute of Physics, Faculty of Natural Sciences and Mathematics, Ss. Cyril
and Methodius University, Arhimedova 3, 1000 Skopje, Macedonia}

\author{\.{I}zzet Sakall\i{}}
\email{izzet.sakalli@emu.edu.tr}
\affiliation{Physics Department, Eastern Mediterranean University, Famagusta, North
Cyprus, via Mersin 10, Turkey}

\author{Ali \"{O}vg\"{u}n}
\email{ali.ovgun@pucv.cl}
\affiliation{Instituto de F\'{\i}sica, Pontificia Universidad Cat\'olica de
Valpara\'{\i}so, Casilla 4950, Valpara\'{\i}so, Chile} 
\affiliation{Physics Department, Eastern Mediterranean University, Famagusta, North Cyprus, via Mersin 10, Turkey}

\date{\today }
\begin{abstract}
In a seminal paper, Alcubierre showed that Einstein's theory of general
relativity appears to allow a super-luminal motion. In the
present study, we use a recent eternal-warp-drive solution found by
Alcubierre to study the effect of Hawking radiation upon an observer located
within the warp drive in the framework of the quantum tunneling method. We
find the same expression for the Hawking temperatures associated with
the tunneling of both massive vector and scalar particles, and show this expression to be
proportional to the velocity of the warp drive. On the other hand, since the
discovery of gravitational waves, the quasinormal modes (QNMs) of black holes
have also been extensively studied. With this puprpose in mind, we perform a QNM
analysis of massive scalar field perturbations in the background of
the eternal-Alcubierre-warp-drive (EAWD) spacetime. Our analytical analysis shows that massive scalar perturbations lead to stable QNMs.
\end{abstract}

\maketitle

\section{Introduction}

General relativity is a playground for many unexpected solutions. One of
them was introduced by M. Alcubierre in 1994, and is known as the EAWD
spacetime \cite{alcub,alcub2}. This solution is found in the original theory
of the general relativity and it allows super-luminal motion by expanding (contracting)
the spacetime behind (in front of) an observer within the spacetime \cite{alcub3}.
It is noted that the observer sits in a locally flat region of the
spacetime, which is the so-called warp bubble. However, the energy-momentum
tensor of EAWD spacetime violates the energy conditions (weak, dominant and strong) in
that weak energy condition requires a negative energy density \cite{alcub4}%
. Today, it is known that negative energy only arises in certain special cases of QFT, such as the
Casimir effect or dark energy in cosmology. On the other hand, the basic premise of EAWD spacetime is widely accepted to have occurred during in the inflationary era of the early universe; when the relative speeds of two co-moving observers in this era are considered, super-luminal motion appears to occur without violating special or general relativity. 

Hiscock was the first to study quantum effects in the EAWD spacetime and showed that the stress--energy diverges if the apparent velocity of the warp bubble (around the spacecraft) exceeds the speed of
light \cite{Hiscock:1997ya}. Finazzi et al. \cite{finazzi}
extended Hiscock's results by investigating semiclassical instability in
dynamical warp drives. Recently, various studies \cite%
{Puthoff:1996my,Pfenning:1997wh,VanDenBroeck:1999sn,GonzalezDiaz:1999db,Clark:1999zn,White:2006iz,GonzalezDiaz:2007zza,Smolyaninov:2010rr,Muller:2011fa,Frauendiener:2011zz}
have appeared in the literature, inspiring us to use the EAWD
metric in our present study.

The discovery of black hole radiation by Hawking and Bekenstein \cite%
{Bekenstein:1993dz,Bekenstein:1975tw,Hawking:1974rv,Bekenstein:1974ax,Bekenstein:1973ur,Bekenstein:1972tm,Bekenstein:1994bc,Hawking:1974sw} showed that black holes are in fact not black, instead they are "gray" and thermally radiate. Today, Hawking radiation is studied in an elegant and widely used way through a quantum tunneling model, among others (see for example \cite%
{Banerjee:2008cf,Jiang:2005ba,Kerner:2006vu,Kerner:2007rr,Kerner:2008qv,Kerner:2007jk,Yale:2008kx,Kruglov1,Kruglov2,Kuang:2017sqa,Ovgun:2017iyb,Ovgun:2017hje,Jusufi:2017vhz,Sakalli:2017ewb,Ovgun:2016roz,Sakalli:2016cbo,Ovgun:2015box,Sakalli:2015jaa,Sakalli:2015taa,GUP0,gup1,gup2,gup3,gup4,gup5,Akhmedova1,tun1,Hawking:2016msc,Akhmedov:2006pg,Akhmedov:2006un,Singleton:2010gz,Singleton:2011vh,Zhu:2009wa,Modak:2012zp,Zampeli:2012tv,Modak:2012cj,Singleton:2013ama,Singleton:2012zz,Modak:2014xza,Akhmedova:2010zz,Akhmedov:2008ru,Birmingham:2001hc,Aros:2002te} and references therein). In the present work, we consider the Hawking radiation of massive scalar and vector particles from the EAWD spacetime. The effect of GUP \cite{gupref} on the Hawking radiation of the EAWD spacetime will also be considered.

QNMs are another important means of probing spacetimes. They represent the characteristic resonance spectrum of a black
hole. By taking the back reaction effects into account, it was shown by
Parikh and Wilczek \cite{Parikh:1999mf} that black holes can radiate energy
with a non-thermal spectrum. Although QNMs dominate in the response of a black
hole to external perturbations, they are affected by Hawking radiation 
\cite{cors1,cors2,cors3,cors4,cors5,cors6}. In the last decade, studies of both QNMs and Hawking radiation have gained momentum (see for instance \cite{mix1,izPRD,mix2,mix3,mix4,mix5,mix6}). In fact, such works
show that the black holes are good testing grounds for quantum gravity
theory \cite%
{izHod,vish,izkg15,Chan:1996yk,izAbram,izFrolov,izAzreg,izNomura,izDaghigh,ortg1,izBecar,aji,Gonzalez:2017shu,Cruz:2015nza,Gonzalez:2012de,Becar:2010zz,Gonzalez:2010vv}%
. Along the same line of thought, in the present study we want to explore
the quantum gravitational outcomes of the EAWD spacetime by considering QNMs in addition to
Hawking radiation.

The Laser-Interferometer-Gravitational-Wave-Observatory (LIGO) has recently made
the `discovery of the century' by detecting the gravitational waves
originating from a merger between two black holes \cite%
{TheLIGOScientific:2016src}. This event once more proves Einstein's
theory of the gravity. QNMs of gravitational waves provide critical
information about the structure of black holes. Thus, QNMs can be a tool to
test general relativity and possible deviations from it \cite{qnmgw,qnmgw2}%
. Moreover, in the AdS/CFT correspondence, QNMs are used to study the
rapidity of a thermal state at the boundary where thermal equilibrium is
established \cite{nunez}. Today, there are many analytical and numerical works on
QNMs in the literature (see for instance\cite%
{Saavedra:2005ug,Lepe:2004kv,Crisostomo:2004hj,Lepe:2012zf,Fernando:2016ftj,Fernando:2015kaa,Fernando:2014gda,Fernando:2012yw,Fernando:2009tv,Fernando:2008hb,Toshmatov:2017qrq,Ferrari:1984zz,Motl:2002hd,Cardoso:2001bb,Konoplya:2003ii,Dreyer:2002vy,Konoplya:2011qq,Birmingham:2001pj,Berti:2009kk,Kokkotas:1999bd,Kuang:2017cgt,Graca:2016cbd,Graca:2015jea,Sakalli:2016jkf,Sakalli:2015uka,Sakalli:2014wja,Sakalli:2011zz,Wang:2000gsa,Wang:2000dt,Wang:2004bv,Wang:2001tk,Birkandan:2017rdp,Jansen:2017oag}%
). To study analytical QNMs, one should solve the field equation on the
considered geometry and derive the one dimensional Schr\"odinger equation or
the so-called Zerilli equation \cite{zerilli} in terms of the tortoise
coordinate $r_{*}$ :

\begin{equation}
{\frac{d^{2}\psi}{dr_{\ast}^{2}}}=[\omega^{2}-V(r_{\ast})]\;\psi,
\end{equation}
where $\omega$ is the frequency of the QNM. In general, $\omega $ %
is labeled by a discrete quantum number ($n=0,1,2,...$)
 and has the following asymptotic form \cite%
{asyfreq,dirtybh}
\begin{equation}
\omega_{n}=\hbox{(offset)}+in\;\hbox{(gap)}+\Game(\frac{1}{\sqrt{n}})\text{
\ \ \ \ as \ \ \ \ }n\rightarrow\infty.
\end{equation}
Here, the \textquotedblleft\hbox{gap}\textquotedblright\ and \textquotedblleft\hbox{offset}\textquotedblright\ are complex parameters
that are determined by the precise form of the spacetime-dependent potential barrier
seen in Eq. (1). In fact, the real part of $\omega $ 
shows the temporal oscillation and its imaginary part describes
the exponential decay.

To compute the QNMs of the EAWD spacetime, we consider the massive Klein-Gordon
equation (KGE). We show that the radial part of the massive KGE reduces to a
hypergeometric differential equation after some manipulation. Thus, we
demonstrate how one can analytically derive the complex QNMs by applying the
appropriate boundary conditions.

The outline of the paper is as follows. In Sec. II, we briefly introduce the
EAWD spacetime. Section III is devoted to calculating the Hawking
temperature of this spacetime using the vector particles' quantum
tunneling with the help of the semi-classical Hamilton--Jacobi method. In
Sec. IV, we consider the quantum tunneling of scalar particles in the
EAWD spacetime. We also study the GUP effect of quantum gravity on the EAWD's
Hawking radiation. In Sec. V, we analytically study the massive
scalar field perturbations in the background of the EAWD spacetime and represent exact
QNMs. We draw our conclusions in Sec. VI.

\section{EAWD Spacetime}

The EAWD spacetime is described by the following line-element, which was
derived by Alcubierre \cite{alcub,alcub2}: 
\begin{equation}
\mathrm{d}s^{2}=-c^{2} \mathrm{d}t^{2}+\left[ \mathrm{d}x-v(r)\mathrm{d}t %
\right] ^{2}+\mathrm{d}x^{2}+\mathrm{d}y^{2}.
\end{equation}
Here, $v(r)$ is the velocity of the spacecraft's moving frame and $r$ represents the distance from the center of the bubble, which is given by 
\begin{equation}
r=\sqrt{(x-v_{0} t)^{2}+y^{2}+z^{2}},
\end{equation}

where $v_{0}$ represents the warp-drive velocity \cite{finazzi}. Furthermore, it is convenient to introduce a new function, $f(r)$, given by $v(r)=v_{0} f(r)$. $f(r)$ should be a smooth function
satisfying the conditions $f(0)=1$ and $f(r) \to0$ if $r\to\infty$. For the
sake of simplicity, in this paper, we shall consider the following $1+1$ dimensional
EAWD metric \cite{finazzi}: 
\begin{equation}
\mathrm{d}s^{2}=-c^{2} \mathrm{d}t^{2}+\left[ \mathrm{d}x-v(r)\mathrm{d}t %
\right] ^{2},
\end{equation}
where $r$ is given as $r=x-v_{0} t$. The choice of $f(r)$ is not unique: we
select the following simple bell-shaped function \cite{finazzi} 
\begin{equation}
f(r)=\frac{1}{\cosh(r/a)}.
\end{equation}

\begin{figure}[h]
\includegraphics[width=0.47\textwidth]{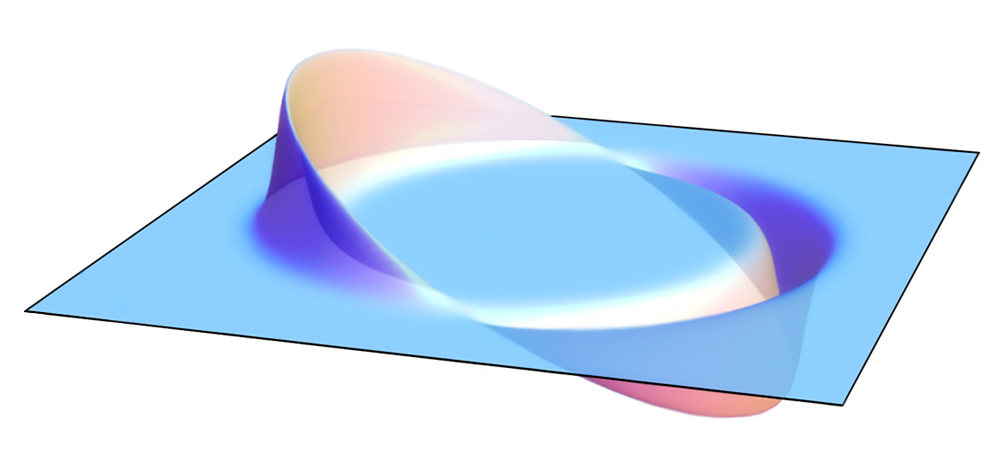}
\caption{{\ Warp field according to the Alcubierre drive drawn \protect\cite%
{alan}}}
\end{figure}

Note that when $v_{0}>c$, we have the super-luminal warp drive. One can write
the above metric as a Painleve metric: 
\begin{equation}  \label{metric1}
\mathrm{d}s^{2}=-c^{2} \mathrm{d}t^{2}+\left[ \mathrm{d}r-\bar{v}(r)\mathrm{d%
}t \right] ^{2},
\end{equation}
where 
\begin{equation}
\bar{v}(r)=v(r)-v_{0}.
\end{equation}

We also note that $\bar{v}(r)<0$ because the warp drive is right going i.e., $%
v(r)\leq v_{0}$, with $v(r)>0$. Introducing $\alpha=v_{0}/c$, the shift
velocity becomes 
\begin{equation}
\bar{v}(r)=\alpha c \left[ \frac{1}{\cosh(r/a)}-1 \right] .
\end{equation}

It is interesting to point out that two horizons appear when $\alpha>1$, by
setting $\bar{v}(r)=-c$. The center of the bubble is located at $r=0$, while
the horizons are located at 
\begin{equation}
r_{h}=r_{1,2}=\mp a \ln\left( \beta+\sqrt{\beta^{2}-1} \right) ,
\end{equation}
with 
\begin{equation}
\beta=\frac{\alpha}{\alpha-1}>1.
\end{equation}

\section{Quantum tunneling of vector particles in EAWD}

The relativistic field equation for a massive vector particles is governed
by the Proca equation (PE) \cite{Kruglov1,Kruglov2} 
\begin{align}  \label{proca}
\frac{1}{\sqrt{-g}}\partial_{\mu}\left( \sqrt{-g}\,\Psi^{\nu\mu }\right) +%
\frac{m^{2}c^{2}}{\hbar^{2}}\Psi^{\nu}=0,
\end{align}
where 
\begin{equation}
\Psi_{\mu\nu}=\partial_{\mu}\Psi_{\nu}-\partial_{\nu}\Psi_{\mu}.  \label{13}
\end{equation}

In our setup, we shall solve the PE in the background of EAWD spacetime %
\eqref{metric1}. The solution is proposed in terms of the WKB approximation
as follows 
\begin{equation}  \label{wkb}
\Psi_{\nu}=C_{\nu}(t,r)\exp\Big(\frac{i}{\hbar}\big(S_{0}(t,r)+\hbar\,
S_{1}(t,r)+\dots\big)\Big).
\end{equation}

Using Eqs. \eqref{proca} and \eqref{wkb}, one finds 
\begin{align}  \label{eq15}
0 & = \left[ \frac{-m^{2}c^{2} \bar{v}(r)+(\partial_{t} S_{0}(t,r))
(\partial_{r} S_{0}(t,r))}{c^{2}}\right] C_{1}-\left[ \frac{(\partial_{r}
S_{0})^{2}+m^{2}c^{2}}{c^{2}}\right] C_{2}, \\
0 & = \left[ \frac{-(\partial_{t} S_{0}(t,r))^{2} +(c^{2}-\bar{v}%
^{2}(r))m^{2}c^{2}}{c^{2}}\right] C_{1}+\left[ \frac{-m^{2}c^{2} \bar {v}%
(r)+(\partial_{t} S_{0}(t,r)) (\partial_{r} S_{0}(t,r))}{c^{2}}\right] C_{2}.
\label{eq16}
\end{align}

By using the symmetries of the spacetime \eqref{metric1}, one may choose the
action as 
\begin{equation}
S_{0}(t,r)=-Et+R(r),
\end{equation}
then by making a substitution from Eqs. \eqref{eq15} and \eqref{eq16}, we
find out 
\begin{widetext}
\begin{eqnarray}
\left[\frac{-m^2 c^2\bar{v}(r)-E R'(r)}{c^2}\right]C_1-\left[\frac{(R'(r))^2+m^2c^2}{c^2}\right]C_2  &=& 0,\\
\left[\frac{-E^2 +(c^2-\bar{v}^2(r))m^2c^2}{c^2}\right]C_1+\left[\frac{-m^2c^2 \bar{v}(r)-E R'(r)}{c^2}\right]C_2 &=& 0.
\end{eqnarray}
\end{widetext}

The physical meaning from these equations can be revealed after we consider
a $2\times2$ matrix equation. In particular, we can choose a matrix $\Xi$
and multiply it with a transpose of a vector $(C_{1}, C_{2}) $, yielding the
matrix equation 
\begin{equation}
\Xi\,(C_{1}, C_{2})^{T}=0.
\end{equation}

This matrix has the following non--zero elements 
\begin{align}
\Xi _{11}& =\Xi _{22}=\frac{-m^{2}c^{2}\bar{v}(r)-ER^{\prime }(r)}{c^{2}}, \\
\Xi _{12}& =-\frac{R^{\prime 2}+m^{2}c^{2}}{c^{2}}, \\
\Xi _{21}& =\frac{-E^{2}+(c^{2}-\bar{v}^{2}(r))m^{2}c^{2}}{c^{2}}.
\end{align}

In order to solve for the radial solution we need to consider $\det \Xi =0$,
which leads to the following differential equation 
\begin{equation}
-{\frac{{m}^{2}\left( \left( -{c}^{2}+\left( \bar{v}\left( r\right) \right)
^{2}\right) R^{\prime 2}-2\,E\bar{v}\left( r\right) R^{\prime }(r)-{c}^{4}{m}%
^{2}+{E}^{2}\right) }{{c}^{2}}}=0.
\end{equation}

This equation can be easily solved for the radial part 
\begin{equation}
R_{\mp}=\int{\frac{E\bar{v} \left( r \right) \mp\sqrt{ {c}^{4}{m}^{2}\left( 
\bar{v}(r)-{c}^{2}\right) +{E}^{2}{c}^{2}}}{ \left( \bar{v}(r)-c\right)
\left( \bar{v}(r)+c \right) }}\mathrm{d}r.
\end{equation}

This integral is singular at the horizon $\bar{v}(r_{h}=r_{1,2})=-c$. Thus,
by expanding in series the velocity near the horizon we find 
\begin{equation}
\bar{v}(r_{h})=-c+\kappa_{1,2}\left( r-r_{1,2}\right) +\mathcal{O}\left[
\left( r-r_{1,2}\right) ^{2} \right],
\end{equation}
where we have defined the surface gravity as 
\begin{equation}
\kappa_{1,2}=\frac{\mathrm{d}v(r)}{\mathrm{d}r}|_{r_{h}=r_{1,2}}=\pm \frac{%
c(\alpha-1)\sqrt{\beta^{2}-1}}{a \beta}\equiv\pm\kappa.
\end{equation}

It is not difficult so see that we find non-zero contribution only for the
ingoing particles moving from the outside to the inside of the EAWD 
\begin{equation}
R_{-}(r_{h}=r_{1,2})=\int\frac{E}{\kappa_{1,2}}\frac{\mathrm{d}r}{(r-r_{1,2})%
}.
\end{equation}

As was noted in \cite{finazzi}, the surface gravity associated with the
first horizon is positive $\kappa_{1}=\kappa>0$, while the one associated
with the second horizon is negative $\kappa_{2}=-\kappa<0$. The physical significance of these two horizons is that they represent a black
and a white horizon, respectively. As was pointed out in Ref. \cite{finazzi}, the choose of the horizon does not lead to a different Hawking temperature in absolute value, therefore we may consider the two surface gravities to
have the same absolute value i.e. $\kappa=|\kappa_{1,2}|$. It is worth noting that `in $3+1$ EAWD spacetime', while one side of the spacetime is expanding (white horizon), the other side of the spacetime contracts (black horizon). In other words, when the particles tunnel from outside to inside near the black horizon, the particles in the vicinity of the white horizon conversely tunnel from inside to outside.  The tunneling
rate is related to the imaginary part of the action in the classically
forbidden region, which is given by 
\begin{equation}
\Gamma\sim e^{-2\,\text{Im}\, S}.
\end{equation}

Therefore, we can use the following identity 
\begin{equation}
\lim_{\epsilon\to0}\text{Im}\frac{1}{r-r_{h}\pm i\epsilon}%
=\pi\delta(r-r_{h}),
\end{equation}
and find a non-zero contribution only for the ingoing radial part $R_{-}(r)$:
\begin{equation}
\text{Im}R_{-}(r)= \frac{ \pi E}{\kappa},\,\,\,\,\text{Im}R_{+}(r)=0.
\end{equation}

We can define the tunneling probability from outside to inside the warp
drive by the following relation 
\begin{equation}
\Gamma=\frac{\exp\left( -2\text{Im} R_{-}\right) }{\exp\left( -2 \text{Im}
R_{+}\right) }=\exp\left( -\frac{2 \pi E}{\kappa}\right) .
\end{equation}

Comparing the above relation with the Boltzmann equation $%
\Gamma_{B}=\exp(-E/T)$, we get the Hawking temperature as follows 
\begin{equation}
T_{H}=\frac{\kappa}{2 \pi}=v_{0} \frac{f^{\prime}(r_{h})}{2\pi}.
\end{equation}

Equation (33) shows that an observer inside the warp drive
experiences a thermal flux of Hawking quanta. In fact, this phenomenon is different from the black hole radiation in which the Hawking quanta
tunnel from inside to outside of the horizon. However, the right way to
recover the above result for the Hawking temperature is to consider the
invariance of canonical transformations \cite%
{Akhmedova:2010zz,Akhmedov:2008ru,Akhmedova1,Akhmedov:2006pg,Akhmedov:2006un}%
. Namely, by considering a closed path, which goes from outside, $r=r_{i}$,
(i.e, just outside of the horizon) to $r=r_{f}$ (just inside of the
horizon): 
\begin{equation}
\oint p_{r}\mathrm{d}r=\int_{r_{i}}^{r_{f}} p_{r}^{in}\mathrm{d}%
r+\int_{r_{f}}^{r_{i}} p_{r}^{out}\mathrm{d}r,
\end{equation}

the total quantum tunneling rate is found to be a sum of the spatial and
temporal contributions. To see this, we draw attention to the pole located
at the horizon $r=r_{h}=r_{1,2}$ seen in Eq. (25) with $\bar{v}(r_{h})=-c$.
But, we can shift the pole by using the Feynman prescription i.e., $r_{h}\to
r_{h}+i\epsilon$:
\begin{equation}
\text{Im} \oint p_{r} \mathrm{d}r= \lim_{\epsilon\to0}\left\lbrace \text{Im}%
\oint\frac{E\bar{v}({r_{h}}) \mp\sqrt{ {c}^{4}{m}^{2}(\bar{v}^{2}(r_{h})-{c}%
^{2})+{E}^{2}{c}^{2}}}{ (\bar{v}(r_{h})-c)\kappa(r_{h}) (r-r_{h} \pm i
\epsilon)} \mathrm{d}r\right\rbrace ,
\end{equation}
in which the relation $p_{r}=\partial_{r}R$ is used. Furthermore, the above
equation can also be written as 
\begin{align}
\text{Im} \oint p_{r} \mathrm{d}r & = \lim_{\epsilon\to0}\Big[\text{Im}%
\int_{r_{i}}^{r_{f}} \frac{E\bar{v}({r_{h}}) + \sqrt{ {c}^{4}{m}^{2}(\bar {v}%
^{2}(r_{h})-{c}^{2})+{E}^{2}{c}^{2}}}{ (\bar{v}(r_{h})-c)\kappa(r_{h})
(r-r_{h} \pm i \epsilon)}\mathrm{d}r\Big]  \notag \\
& +\lim_{\epsilon\to0}\Big[ \text{Im} \int_{r_{f}}^{r_{i}} \frac{E\bar {v}({%
r_{h}}) - \sqrt{ {c}^{4}{m}^{2}(\bar{v}^{2}(r_{h})-{c}^{2})+{E}^{2}{c}^{2}}}{
(\bar{v}(r_{h})-c)\kappa(r_{h}) (r-r_{h} \pm i \epsilon)}\mathrm{d}r \Big].
\end{align}

One can see from the last equation that there is no contribution to the
imaginary part coming from the first term. This is due to the Painleve
coordinates which means that the particle experiences barrier only from
outside the horizon to inside (not from the reverse way). Hence, the spatial
contribution reads 
\begin{equation}
\text{Im}\oint p_{r}\mathrm{d}r=\frac{\pi E}{\kappa }.
\end{equation}

We shall now proceed to find the temporal contribution. To this end, we express the EAWD
spacetime in the following compact form (Painleve coordinates):
\begin{equation}
\mathrm{d}s^{2}=-\Big(c^{2}-\bar{v}^{2}(r)\Big)\mathrm{d}\tilde{t}^{2}+\frac{%
c^{2}}{c^{2}-\bar{v}^{2}(r)}\mathrm{d}r^{2},  \label{ADW}
\end{equation}
under the coordinate transformations 
\begin{equation}
\mathrm{d}t=\mathrm{d}\tilde{t}-\frac{\bar{v}(r)}{c^{2}-\bar{v}^{2}(r)}%
\mathrm{d}r,
\end{equation}
where $t$ is the Painleve time. Using the action Eq. (17), we
find 
\begin{equation}
S_{0}=-E\tilde{t}+\int\frac{\bar{v}(r)\,E}{c^{2}-\bar{v}^{2}(r)}\mathrm{d}%
r+R(r)
\end{equation}

By solving the integral, we can find the temporal contribution as 
\begin{equation}
\text{Im}(E\Delta t^{out,in})=\frac{\pi E}{2 \kappa}.
\end{equation}

Following the paper of Akhmedova et al \cite{Akhmedova1}, the total
tunneling rate is obtained as follows 
\begin{align}
\Gamma & =\exp\left[ -\left( \text{Im} (E\Delta t^{out})+\text{Im}(E\Delta
t^{in})+\text{Im }\oint p_{r} \mathrm{d}r\right) \right]  \notag \\
& = \exp\left( - \frac{2 \pi E}{\kappa}\right) .
\end{align}

Employing the Boltzmann formula $\Gamma_{B}=e^{-E/T_{H}}$, we find the
foreknown Hawking temperature: 
\begin{equation}
T_{H}=\frac{\kappa}{2 \pi}=v_{0} \frac{f^{\prime}(r_{h})}{2\pi}.
\end{equation}

\section{Quantum tunneling of scalar particles and the effect of GUP on Hawking radiation}

In this section, we shall focus on the Hawking temperature by using the
tunneling of the scalar particle from the EAWD spacetime. We start with the
KGE 
\begin{equation}
\frac{1}{\sqrt{-g}}\partial_{\mu}\left( \sqrt{-g}g^{\mu\nu}\partial_{\nu}
\Phi\right) -\frac{m^{2}c^{2}}{\hbar^{2}}\Phi=0,  \label{iz1}
\end{equation}
on the background of the metric (38), which can be rewritten as 
\begin{equation}
\mathrm{d}s^{2}=-F\mathrm{d}\tilde{t}^{2}+\frac{1}{G}\mathrm{d}r^{2},
\end{equation}
where $F=c^{2}-\bar{v}^{2}(r)$ and $G=\frac{c^{2}-\bar{v}^{2}(r)}{c^{2}}$.
From those two equations, we obtain 
\begin{equation}
-\frac{1}{F}\partial_{t}^{2} \Phi+G\,\partial_{r}^{2} \Phi+\frac{1}{2}\frac {%
G}{F}F^{\prime} \partial_{r}\Phi+\frac{1}{2}G^{\prime} \partial_{r}\Phi -%
\frac{m^{2} c^{2}}{\hbar^{2}} \Phi=0.
\end{equation}

Then, we make use of the WKB ansatz: 
\begin{equation}
\Phi=\exp\left( {\frac{i}{\hbar} S(r,t)}\right) .
\end{equation}
In the limit of $\hbar$ goes to zero, we obtain the relativistic
Hamilton--Jacobi equation: 
\begin{equation}
\frac{1}{F}(\partial_{t} S)^{2}-G(\partial_{r} S)^{2}-m^{2} c^{2} =0.
\end{equation}

We now suppose the solution of the action as the following 
\begin{equation}
S(r,t)=-\omega t+W(r).
\end{equation}
The radial part $W(r)$ is obtained from 
\begin{equation}
W(r)=\pm\int\frac{\mathrm{d}r}{\sqrt{FG}}\sqrt{E^{2}-m^{2}c^{2}F}.
\end{equation}

The above integral is nothing but the 
\begin{equation}
W(r)=\pm\int\frac{c\,\sqrt{E^{2}-m^{2}c^{2}F}}{(c-\bar{v}(r))(c+\bar{v}(r))}%
\mathrm{d}r.
\end{equation}

Note that there is a pole at the horizon $F(r_{h})=0$. After using the near
horizon approximation and residue theorem, we solve the complex integral and
find the following solution: 
\begin{equation}
W(r_{h})=\pm\frac{\pi i E}{2\,\kappa(r_{h})}+\text{real contribution.}
\end{equation}

The spatial contribution to the tunneling can be calculated as 
\begin{align}
\Gamma_{spatial} & \propto\exp\left( -\text{Im} \oint p_{r} \mathrm{d}%
r\right)  \notag \\
& =\exp\left[ -\text{Im} \left( \int p_{r}^{+}\mathrm{d}r-\int p_{r}^{-}%
\mathrm{d}r\right) \right]  \notag \\
& = \exp\left( -\frac{ \pi E}{ \kappa(r_{h})}\right) .
\end{align}

The temporal part contribution is revealed after we consider the connection
of the interior region and the exterior region of the EAWD spacetime.
Introducing $t\to t -i\pi/(2\kappa) $, we will have Im ($E\Delta
t^{out,in})=E\pi /(2\kappa)$. Then the total temporal contribution for a
round trip gives 
\begin{align}
\Gamma_{temporal} & \propto\exp\left[ -\left( \text{Im}(E\Delta t^{out})+%
\text{Im}(E\Delta t^{in}\right) \right]  \notag \\
& =\exp\left( -\frac{ \pi E}{ \kappa(r_{h})}\right) .
\end{align}

The total tunneling rate of the particles tunneling from outside to the
inside reads 
\begin{align}
\Gamma & = \exp\Big[-\Big(\text{Im}(E\Delta t^{out})+\text{Im}(E\Delta
t^{in})+\text{Im}\oint p_{r} \mathrm{d}r \Big)\Big]= \exp\left[ - \frac{2
\pi E}{ \kappa(r_{h})}\right] .  \notag
\end{align}

Hence the Hawking temperature is obtained as 
\begin{equation}
T_{H}=\frac{\kappa}{2 \pi}=v_{0} \frac{f^{\prime}(r_{h})}{2\pi},
\end{equation}
in full agreement to previous result. Let us now consider the GUP effect on
the scalar particle tunneling the EAWD black hole using the modified
commutation relations and modification of the KGE with GUP 
\cite{GUP0,gup1,gup2,gup3,gup4,gup5}, $\Phi$ as follows: 
\begin{equation}
-(i\hslash)^{2}\partial^{t}\partial_{t} \Phi=\left[ (i\hslash)^{2}\partial
^{i}\partial_{i}+m_{p}^{2}c^{2}\right] \left[ 1-2\alpha_{GUP}\left(
(i\hslash)^{2}\partial^{i}\partial_{i}+m_{p}^{2}c^{2}\right) \right] \Phi,
\label{11}
\end{equation}
where $\alpha_{GUP}$\ is the GUP parameter, and $m_{p}$ is the mass of the
scalar particle. After using WKB ansatz, we find the
following equation in the leading order of $\hbar$ as follows:

\begin{equation}
\frac{1}{F}(\partial_{t}\mathcal{S})^{2}=G\,(\partial_{r}\mathcal{S}%
)^{2}+m_{p}^{2}c^{2}\left( 1-2\,\alpha_{GUP}\,G(\partial_{r}\mathcal{S}%
)^{2}-2\alpha_{GUP}m_{p}^{2}c^{2}\right) .  \label{133}
\end{equation}

Then, we impose the Hamilton--Jacobi ansatz: 
\begin{equation}
\mathcal{S}(t,r )=-E\,t+W(r),  \label{14}
\end{equation}

and thus Eq. (57) reduces to

\begin{equation}
\frac{1}{F}E^{2}=G\,(W^{\prime})^{2}+m_{p}^{2}c^{2}\left( 1-2\alpha
_{GUP}G(W^{\prime})^{2}-2\alpha_{GUP}m_{p}^{2}c^{2}\right) .  \label{15}
\end{equation}

For the above equation, we obtain the outgoing and ingoing ($\pm$) radial
solutions 
\begin{widetext}
\begin{equation}
W(r)_{\pm}=\pm \int \frac{1}{\sqrt{F G}}\frac{\sqrt{E^{2}-F\left(m_{p}^{2}c^2\right)
\left( 1-2m_{p}^{2}c^2\alpha _{GUP}\right) }}{\sqrt{1-2m_{p}^{2}c^2\alpha _{GUP}}}%
\mathrm{d}r.  \label{16}
\end{equation}%
\end{widetext}

As the integral becomes zero at the horizon, we use the complex-integral path
method to find the solution near the horizon. Namely, we get 
\begin{equation}
T_{GUP}=\frac{\kappa}{2 \pi}\sqrt{1-2m_{p}^{2}c^{2}\alpha_{GUP}}=v_{0} \frac{%
f^{\prime}(r_{h})}{2 \pi}\sqrt{1-2m_{p}^{2}c^{2}\alpha_{GUP}}.  \label{25}
\end{equation}

If we ignore the GUP effect, i.e., $\alpha_{GUP}=0$, the original Hawking
temperature is recovered. Let us note that a similar analysis can be applied to the GUP effects of vector particles, yielding a similar conclusion. Moreover, in Ref. \cite{Li}, it was shown that the GUP Hawking temperature related to the vector particles is also affected by the nature of particles (i.e. their mass and spin), such that the GUP Hawking temperature would be slightly different from that related to scalar particles. The latter remark might be important for Planck-scale physics. In our case, however, we work in the $1+1$ EAWD metric; hence, the result will be quite similar. Therefore, in order to see the difference, one needs to consider the problem of quantum tunneling from the $3+1$ EAWD spacetime.

\section{QNMs of particular EAWD spacetime}

In this section, we explore the analytical forms of the QNMs of the
(1+1)-dimensional EAWD spacetime given as Eq. (\ref{ADW}): 
\begin{equation}
\mathrm{d}s^{2}=-N\mathrm{d}\tilde{t}^{2}+\frac{c^{2}}{N}\mathrm{d}r^{2},
\label{iz0}
\end{equation}
where $N=c^{2}-\bar{v}^{2}(r).$ Without loss of generality, throughout this
section, we use geometric unit system ($c=\hbar=1$) and focus on a
particular position-dependent velocity: $\bar{v}(r)\equiv\overline{f}(r)=e^{-%
\frac{r}{2}}$ [a suitable smooth function satisfying the conditions $\overline{f}(0)=1$ and $\overline{f}(r\rightarrow\infty)\rightarrow0$], 
 which represents a velocity that exponentially decreases with $r$%
. We want to stress that all results in this section depend on
this particular choice of $\bar{v}(r)$. Because, according to our
observations, this is the probably only the case for which one can
perform a complete analytical computation of QNMs in the EAWD spacetime.
For a discussion of the velocity function of EAWD spacetime, the reader is referred to 
\cite{alcub4} and references therein.

We first consider the massive KGE: 
\begin{equation}
\frac{1}{\sqrt{-g}}\partial_{\alpha}\left( \sqrt{-g}g^{\alpha\nu}\partial_{%
\nu}\varphi\right) -\mu^{2}\varphi=0,  \label{iz111}
\end{equation}

where $\mu$ is the mass of the scalar field $\varphi$. Choosing the
following ansatz for the scalar field 
\begin{equation}
\varphi=e^{-i\omega t}\mathcal{R}(r),  \label{iz2}
\end{equation}
where $\omega$ is the frequency or energy of the flux of scalar particles at
spatial infinity, Eq. (\ref{iz111}) is shown to be separable in the EAWD
background (\ref{ADW}). Then, one can reduce the massive KGE to the
following radial equation 
\begin{equation}
N\partial_{r}^{2}\mathcal{R}(r)+e^{-r}\partial_{r}\mathcal{R}(r)+{\left( 
\frac{\omega^{2}-\mu^2N}{N}\right) \mathcal{R}(r)=0}.  \label{iz3}
\end{equation}
After setting a new variable {$z=N=1-e^{-r}$}, Eq. (\ref{iz3}) becomes

\begin{equation}
z\left( 1-z\right) \partial_{z}^{2}\mathcal{R}\left( z\right) +\left(
1-2z\right) \partial_{z}\mathcal{R}\left( z\right) +{\frac{{\omega}^{2}-{\mu}%
^{2}z}{z\left( 1-z\right) }}\mathcal{R}\left( z\right) =0.  \label{iz4}
\end{equation}

In sequel, if one uses the following s-homotopic transformation \cite%
{izkg15,izPRD,ortg1}

\begin{equation}
\mathcal{R}(z)=z^{\alpha}(1-z)^{\beta}\mathcal{F}(z),  \label{iz5}
\end{equation}

where

\begin{equation}
\alpha=-i\omega,  \label{iz6}
\end{equation}

\begin{equation}
\beta=i\sqrt{\omega^{2}-{\mu}^{2}},  \label{iz7}
\end{equation}

the radial equation (\ref{iz4}) turns out to be a hypergeometric
differential equation \cite{izAbram}: 
\begin{equation}
z(1-z)\partial_{z}^{2}\mathcal{F}(z)+\left[ \widehat{c}-(\widehat{a}+%
\widehat{b}+1)z\right] \partial_{z}\mathcal{F}(z)-\widehat{a}\widehat{b}%
\mathcal{F}(z)=0.  \label{iz8}
\end{equation}

The coefficients $\widehat{a},$ $\widehat{b},$ and $\widehat{c}$ are given by

\begin{equation}
\widehat{a}=\alpha+\beta=i\omega\left( \sqrt{1-\frac{{\mu}^{2}}{\omega^{2}}}%
-1\right) ,  \label{iz9}
\end{equation}

\begin{equation}
\widehat{b}=\alpha+\beta+1=i\omega\left( \sqrt{1-\frac{{\mu}^{2}}{\omega^{2}}%
}-1\right) +1,  \label{iz10}
\end{equation}

\begin{equation}
\widehat{c}=1+2\alpha=1-2i\omega.  \label{iz11}
\end{equation}

The three regular singular points of Eq. (\ref{iz8}) are located at $z=0,$ $%
z=1,$ and $z=\infty$. There are two independent solutions of Eq. (\ref{iz8}) 
\cite{izAbram}: 
\begin{equation}
\mathcal{F}(z)=A_{1}F(\widehat{a},\widehat{b};\widehat{c};z)+A_{2}z^{1-\widehat{c}}F(1+%
\widehat{a}-\widehat{c}+1,1+\widehat{b}-c;2-\widehat{c};z),  \label{iz12}
\end{equation}
where $A_{1},A_{2}$ are constants and $F(\widehat{a},\widehat{b};\widehat{c}%
;z)$ stands for the Gaussian hypergeometric function \cite{izAbram}. Thus,
the analytical solution of Eq. (\ref{iz4}) is given by 
\begin{equation}
\mathcal{R}(z)=A_{1}z^{-i\omega}(1-z)^{i\sqrt{\omega^{2}-{\mu}^{2}}}F(\widehat{a},%
\widehat{b};\widehat{c};z)+  \notag \\
A_{2}z^{i\omega}(1-z)^{i\sqrt{\omega^{2}-{\mu}^{2}}}F(1+\widehat{a}-\widehat{%
c}+1,1+\widehat{b}-c;2-\widehat{c};z).  \label{iz13n}
\end{equation}

Meanwhile, it is worth noting that setting\ $\omega=0,$ which corresponds
to\ $\widehat{c}=1$, two solutions of Eq. (\ref{iz12}) become linearly
dependent. In this case, the general solution represents a bound state Ref. 
\cite{izFrolov}.

In the vicinity of the event horizon $(z\rightarrow0)$, the radial function $%
\mathcal{R}\left( z\right) $ behaves as 
\begin{equation}
\mathcal{R}\left( z\right) \sim A_{1}e^{-i\omega\ln z}+A_{2}e^{i\omega\ln z}.
\label{iz14}
\end{equation}
Namely, the near horizon form of the scalar field $\varphi$ reads 
\begin{equation}
\varphi\sim A_{1}e^{-i\omega(t+\ln z)}+A_{2}e^{-i\omega(t-\ln z)}.
\label{iz15}
\end{equation}

It is clear from Eq. (\ref{iz15}) that the first term represents the ingoing
wave while the second term stands for the outgoing wave. For obtaining the
QNMs, one should impose the requirement that there exist only ingoing waves
at the event horizon. Therefore, to satisfy this condition, we set $A_{2}$ $%
=0$. Thus, the physically acceptable radial solution for QNMs is given by

\begin{equation}
\mathcal{R}\left( z\right) =A_{1}z^{-i\omega}(1-z)^{i\omega\sqrt{1-\frac{{\mu}^{2}}{%
\omega^{2}}}}F(\widehat{a},\widehat{b};\widehat{c};z).  \label{iz16}
\end{equation}

For matching the near horizon and asymptotic regions, we are interested in
the large $r$ behavior ($z\rightarrow1$) of the solution (\ref{iz16}). To
this end, we use the linear transformation law $z\rightarrow1-z$ for the
hypergeometric functions \cite{izAbram}: 
\begin{align}
\mathcal{R}\left( z\right) & =A_{1}z^{-i\omega}(1-z)^{i\omega\sqrt{1-\frac{{\mu}^{2}}{%
\omega^{2}}}}\frac{\Gamma(\widehat{c})\Gamma(\widehat{c}-\widehat{a}-%
\widehat{b})}{\Gamma(\widehat{c}-\widehat{a})\Gamma(\widehat{c}-\widehat{b})}%
F(\widehat{a},\widehat{b};\widehat{a}+\widehat{b}-\widehat{c}+1;1-z)+  \notag
\\
& +A_{1}z^{-i\omega}(1-z)^{-i\omega\sqrt{1-\frac{{\mu}^{2}}{\omega^{2}}}}%
\frac{\Gamma(\widehat{c})\Gamma(\widehat{a}+\widehat{b}-\widehat{c})}{\Gamma(%
\widehat{a})\Gamma(\widehat{b})}F(c-\widehat{a},\widehat{c}-\widehat{b};%
\widehat{c}-\widehat{a}-\widehat{b}+1;1-z).  \label{iz17}
\end{align}

Therefore, the asymptotic form (around $z=1$) of the radial solution (\ref%
{iz16}) is given by 
\begin{equation}
\mathcal{R}\left( z\right) \sim A_{1}(1-z)^{i\omega\sqrt{1-\frac{{\mu}^{2}}{\omega^{2}}}%
}\frac{\Gamma(\widehat{c})\Gamma(\widehat{c}-\widehat{a}-\widehat{b})}{%
\Gamma(\widehat{c}-\widehat{a})\Gamma(\widehat{c}-\widehat{b})}%
+A_{1}(1-z)^{-i\omega\sqrt{1-\frac{{\mu}^{2}}{\omega^{2}}}}\frac {\Gamma(%
\widehat{c})\Gamma(\widehat{a}+\widehat{b}-\widehat{c})}{\Gamma(\widehat{a}%
)\Gamma(\widehat{b})}.  \label{iz18}
\end{equation}

Correspondingly, the near spatial infinity form of the scalar field becomes

\begin{equation}
\varphi\sim A_{1}e^{-i\omega\left[ t-\sqrt{1-\frac{{\mu}^{2}}{\omega^{2}}}%
\ln(1-z)\right] }\frac{\Gamma(\widehat{c})\Gamma(\widehat{c}-\widehat{a}-%
\widehat{b})}{\Gamma(\widehat{c}-\widehat{a})\Gamma(\widehat{c}-\widehat{b})}%
+A_{1}e^{-i\omega\left[ t+\sqrt{1-\frac{{\mu}^{2}}{\omega^{2}}}\ln(1-z)%
\right] }\frac{\Gamma(\widehat{c})\Gamma(\widehat{a}+\widehat{b}-\widehat{c})%
}{\Gamma(\widehat{a})\Gamma(\widehat{b})}.  \label{iz19}
\end{equation}

QNMs impose another requirement that the ingoing waves spontaneously must
die out at spatial infinity, which means that only outgoing waves are
allowed to survive at the infinity. To distinguish the outgoing and ingoing
waves in Eq. (\ref{iz19}), we impose the following condition

\begin{equation}
{\sqrt{1-\frac{{\mu}^{2}}{\omega^{2}}}\in%
\mathbb{R}
>0}.  \label{izcon}
\end{equation}

Thus, $A_{1}$ and $A_{2}$ become the amplitudes
of inoing and outgoing waves, respectively. However, the boundary conditions
of QNM impose that the first term of Eq. (\ref{iz19}) must be terminated.
This can be performed by the poles of the gamma functions [$\Gamma(\widehat{c%
}-\widehat{a})$ or $\Gamma(\widehat{c}-\widehat{b})$] seen in the
denominator of the first term of Eq. (\ref{iz19}). As is well-known, the
gamma functions $\Gamma(y)$ have the poles at $y=-n$ for $n=0,1,2,...$. So,
the massive scalar waves of the EAWD spacetime's QNMs impose the following
restrictions: 
\begin{equation}
\widehat{c}-\widehat{a}=-n,\text{ \ \ or \ \ }\widehat{c}-\widehat{b}=-n,%
\text{ \ \ \ \ \ \ \ \ (}n=0,1,2,...\text{)}  \label{iz20}
\end{equation}
From above, we get two sets of QNMs: 
\begin{equation}
\omega_{\text{set1}}=-\frac{i}{2}\left( n-\frac{{\mu}^{2}}{n}\right) ,
\label{iz21}
\end{equation}

and

\begin{equation}
\omega_{\text{set2}}=-\frac{i}{2}\left( 1+n-\frac{{\mu}^{2}}{1+n}\right) .
\label{iz22}
\end{equation}

It is worth noting that for the set of Eq. \ref{iz21}, in order to
avoid the divergence of the frequency, one should exclude $n=0$ case and consider $n=1,2,3,..$. Similar results were obtained in
the QNMs of spin-$\frac{1}{2}$ waves propagating in the geometry
of Witten black hole \cite{Witten}. Having stable QNMs, one must have $%
\text{Im}\omega <0$. Therefore, when we analyze the obtained QNMs,
it can be seen that the first set (\ref{iz21}) admits the stable modes when

\begin{equation}
{\mu}<n\text{ \ \ \ \ \ \ \ }\left( n\geq1\right) ,  \label{set1}
\end{equation}

and the second set (\ref{iz22}) is stable if

\begin{equation}
{\mu}<n+1\text{ \ \ \ \ \ \ \ \ }\left( n\geq0\right) .  \label{set2}
\end{equation}

On the other hand, one may question the existence of the unstable
modes depending on the value of $\mu$. However, we want to draw the reader's
attention to Eq. (\ref{izcon}), which allows us to decide the type of wave:
ingoing or outgoing. When the obtained sets (\ref{iz21}) and (\ref{iz22})
[with stability requirements given by: Eqs. (\ref{set1}) and (\ref{set2})] are
used in Eq. (\ref{izcon}), we thus have

\begin{equation}
\left. \sqrt{1-\frac{{\mu}^{2}}{\omega^{2}}}\right\vert _{\omega =\omega_{%
\text{set1}}}=\frac{n^{2}+{\mu}^{2}}{n^{2}-{\mu}^{2}}>0,\text{ \ \ \ (for }{%
\mu}<n\text{ and\ }n\geq1\text{),}  \label{stab1}
\end{equation}

and

\begin{equation}
\left. \sqrt{1-\frac{{\mu}^{2}}{\omega^{2}}}\right\vert _{\omega =\omega_{%
\text{set2}}}=\frac{\left( n+1\right) ^{2}+{\mu}^{2}}{\left( n+1\right) ^{2}-%
{\mu}^{2}}>0,\text{ \ \ \ (for }{\mu}<n+1\text{ and\ }n\geq0\text{).}
\label{stab2}
\end{equation}

One can easily deduce from Eqs. (\ref{stab1}) and (\ref{stab2}) that
only stable QNMs can fulfill the wave-identifier condition (\ref{izcon}).
Namely, QNMs that do not satisfy conditions Eqs. (\ref{set1}) and (\ref%
{set2}) for being stable will also not satisfy Eq. (\ref{izcon}), which allows
us to recognize ingoing and outgoing waves at spatial infinity. In short,
the analytical method that we followed encompasses only the stable QNMs. On
the other hand, our findings must not be interpreted to mean that there are no unstable waves
in the EAWD spacetime; instead they must be studied (including the
numerical methods) in more detail. As a matter of fact, it is known that in
some geometries there are unstable QNMs (see for instance \cite%
{izBecar,izAzreg}). Moreover, in the limit of highly damped modes (i.e., $%
n\rightarrow\infty$) the QNMs of the EAWD spacetime become mass independent. The
latter remark is also in good agreement with previous studies \cite%
{izNomura,izDaghigh,izHod,izBecar}.

\section{Conclusions}

In this paper, we have studied the Hawking radiation of the EAWD spacetime in terms of 
the quantum tunneling method. Our results obtained from the quantum tunneling
of massive vector and scalar particles are in accordance with the
statistical Hawking temperature, which is simply equal to $\frac{\kappa}{2\pi%
}$. We have also shown that the Hawking temperature can be decreased when the
GUP effects are taken into account. QNM analysis of the particular EAWD
spacetime (\ref{iz0}) has shown that when this spacetime is perturbed by the
massive scalar particles, QNMs based on the wave-identifier
condition (\ref{izcon})\ satisfy the stability conditions (\ref{iz21}) and (%
\ref{iz22}). Namely, for the exact solution (\ref{iz13n}) of the KGE we
obtained, the unstable modes do not fulfill Eq. (\ref{izcon}): therefore
only stable QNMs are taken into account. A similar result was obtained for the
QNM frequencies of the Dirac field propagating in the uncharged Witten black
hole \cite{Witten}.

In future work, we will extend our analysis to the problem of the
greybody factor in the EAWD spacetime. In this way, we plan to find
analytical expressions for the absorption cross-section, as well as for the
decay-rate for the scalar field in the EAWD spacetime.

\begin{acknowledgments}
We wish to thank the Editor and anonymous Reviewers for their valuable
comments and suggestions. Special thanks to Professor A. L\'{o}pez-Ortega for
helpful discussions and correspondence. This work was supported by the
Chilean FONDECYT Grant No. 3170035 (A\"{O}). A\"{O} is grateful to the CERN theory (CERN-TH) division for their hospitality where part of this work was done.
\end{acknowledgments}

\end{document}